\documentclass[graybox]{svmult}

\usepackage{type1cm}        % activate if the above 3 fonts are
\usepackage{makeidx}         % allows index generation
\usepackage{graphicx}        % standard LaTeX graphics tool
\usepackage{multicol}        % used for the two-column index
\usepackage[bottom]{footmisc}% places footnotes at page bottom

\usepackage{newtxtext}       % 
\usepackage{newtxmath}       % selects Times Roman as basic font

\makeindex             % used for the subject index
\usepackage{url}

\begin{document}
% preprint number: 
% DESY 21-047 \\
% KA-TP-05-2021 

\title*{Calculating Four-Loop Corrections in QCD}
\author{S. Moch and V. Magerya}
\institute{S. Moch \at II. Institut f\"ur Theoretische Physik, Universit\"at Hamburg \\
   Luruper Chaussee 149, D--22761 Hamburg, Germany\\ \email{sven-olaf.moch@desy.de}
\and V. Magerya \at Institut f\"ur Theoretische Physik, Campus S\"ud \\
Karlsruher Institut f\"ur Technologie (KIT), 
D--76128 Karlsruhe, Germany\\ \email{vitalii.maheria@kit.edu}}
\maketitle

\abstract{
We review the current status of perturbative corrections in QCD at four loops 
for scattering processes with space- and time-like kinematics at colliders, 
with specific focus on deep-inelastic scattering and electron-positron annihilation.
The calculations build on the parametric reduction of loop and phase space integrals
up to four-loop order using computer algebra programs such as {\tt Form},
designed for large scale computations.
}

\section{Introduction}
\label{sec:intro}

Perturbation theory forms the backbone of theory predictions for scattering
processes at high energy colliders. 
Given the size of the coupling constant $\alpha_s$ in the theory of strong
interactions, Quantum Chromodynamics (QCD), this requires
the computation of quantum corrections at higher orders.
For collisions involving hadrons, either in the initial state or identified in the final
state, the theory description is based on  QCD factorization, 
which allows for the separation of long- and short-distance physics.
Within this framework, quantum corrections to the hard scattering cross
section driven by short-distance physics are calculated typically at the
next-to-next-to-leading order (NNLO) in order to reach an accuracy of the
order of a few percent from the truncation of the perturbative expansion. 
The long-distance physics part of the interaction is encoded in hadronic matrix elements
which are inaccessible to perturbation theory. 
Based on the description of a hadron as an incoherent combination of parton states,
it is possible, though, to compute matrix elements of partonic operators and,
in particular, their scale dependence in perturbative QCD.
The gained knowledge, encoded in splitting functions, 
serves as input to the parton evolution equations derived 
from the renormalization group and forms an essential ingredient in the 
determination of the non-perturbative parton distribution functions
(PDFs) or parton fragmentation functions (FFs) from fits to experimental data.
The current description of QCD evolution equations for PDFs and FFs is complete at NNLO.
This requires the splitting functions in space-like and time-like kinematics 
at the three-loop level~\cite{Moch:2004pa,Vogt:2004mw,Mitov:2006ic,Moch:2007tx,Almasy:2011eq,Chen:2020uvt} 
as well as the coefficient functions for the hard scattering at two-loop order 
entering, e.g., in DIS structure functions~\cite{vanNeerven:1991nn,Zijlstra:1991qc,Zijlstra:1992qd,Moch:1999eb}
or in fragmentation functions 
in $e^+e^-$ annihilation~\cite{Rijken:1996vr,Rijken:1996ns,Rijken:1996npa,Mitov:2006ic,Mitov:2006wy}.

With the increasing precision of the experimental data collected at the Large
Hadron Collider (LHC) for Standard Model (SM) processes used to extract 
fundamental theory parameters such as the strong coupling $\alpha_s$ or the
PDFs~\cite{Accardi:2016ndt}, the step towards the next-to-next-to-next-to-leading
order (N$^3$LO) becomes necessary.
This is particularly crucial in preparation for the physics
program at a future Electron-Ion Collider (EIC)~\cite{Boer:2011fh,Accardi:2012qut}, 
where PDFs as well as parton FFs are expected to be accessible with high precision,
but also in view of the ongoing future circular collider (FCC) studies~\cite{Blondel:2018mad}.
The push beyond the state-of-the-art requires the calculation of four-loop
corrections, building on known results for the renormalization of QCD
at four-loop~\cite{vanRitbergen:1997va,Czakon:2004bu}
and even five-loop order~\cite{Baikov:2016tgj,Herzog:2017ohr,Luthe:2017ttg}.

The simplest cross section computations at the four-loop level involve 
semi-inclusive (single-scale) observables, 
such as DIS structure functions and $e^+e^-$ fragmentation functions and 
the current status of their calculation will be discussed and reviewed in detail below.

\section{Space-like kinematics}
\label{sec:space}

The scattering reaction for unpolarized DIS reads
\begin{eqnarray}
  \label{eq:dis}
  l(k) \:+\: {\rm nucl}(p) \:\:\rightarrow\:\: l^{\,\prime} (k^{\,\prime}) \:+\: X 
  \:\: ,
\end{eqnarray}
where $l$ and $l^{\,\prime}$ denote the scattered lepton 
and `nucl' a nucleon with respective momenta $k$, $k^{\,\prime}$ and $p$. 
$X$ summarizes the remaining hadronic final states.
The inclusive DIS cross section factorizes as $d \sigma \:\sim\: L^{\,\mu\nu} W_{\mu\nu\,}$ 
in terms of leptonic and hadronic tensors $L_{\mu\nu}$ and $W_{\mu\nu}$.
The latter one encodes the strong interaction dynamics and 
can be expanded to define the unpolarized structure functions $F_{\,2,\:3\:,L}$, 
\begin{eqnarray}
  \label{eq:htensor}
  W_{\mu\nu}(p,q) 
  & = & 
    \frac{1}{4\pi} \int \! d^{\,4}z\; {\rm{e}}^{\,{\rm{i}}q \cdot z} \,
    \langle {\,\rm{nucl,}\,p}\vert J_{\mu}^\dagger(z) J_{\nu}(0)\vert \,
    {\rm{nucl,}\,p}\rangle  \nonumber \\
  & = &
    \frac{e_{\mu\nu}}{2x_B}\: F_{L}(x_B,Q^2) \:+\:
    \frac{d_{\mu\nu}}{2x_B}\: F_{2}(x_B,Q^2) \:+\: 
    {\rm{i}}\,\frac{\epsilon_{\mu\nu p q}}{p\cdot q}\: F_{3}(x_B,Q^2) 
    \:\: .
\end{eqnarray}
Here $J_{\mu}$ represents an electro-magnetic or weak current.
The momentum $q$ is transferred by the gauge-boson 
with space-like kinematics, $Q^2 \equiv -q^2 > 0$, 
and the Bjorken variable is defined as 
\begin{eqnarray}
  \label{eq:xb}
  x_B\,=\, \frac{Q^2}{2\,p\cdot q}\, ,
\end{eqnarray}
with $0 < x_B \leq 1$. 
The symmetric tensors $e_{\mu\nu}$ and $d_{\mu\nu}$ multiplying the structure
functions $F_{\,2,\:L}$ are dependent on $p$ and $q$, 
while the totally antisymmetric one $\epsilon_{\mu\nu\alpha\beta}$ 
in front of the structure function $F_3$ 
arises from the vector$/$axial-vector interference, 
see~\cite{Vermaseren:2005qc,Moch:2008fj} for definitions.

QCD factorization allows for the decomposition of the DIS structure functions 
in terms of (space-like) coefficient functions $C_{a,{\rm f}}$ and PDFs $\phi_{\rm f}$,
\begin{equation}
\label{eq:FdispQCD}
  F_a(x_B,Q^2) \;\: = \; \sum_{\rm f = q,\,\bar{q},\,g} \;
  \int_{x_B}^1 {dz \over z} \; \phi_{\rm f}\bigg( \,{x_B \over z},\mu^2\bigg)\, 
  C_{a,{\rm f}} \left( z,\alpha_{\rm s}(\mu^2),{\mu^2 \over Q^2} \right)
  \; + \; {\cal O} \!\left( \frac{1}{Q^2} \right) 
  \:\: ,
\end{equation}
up to higher-twist corrections ${\cal O}(1/Q^2)$. 
The coefficient functions can be computed in perturbation theory 
via expansions in the strong coupling $a_{\rm s} \,\equiv\, \alpha_{\rm s}/(4\pi)$ 
as 
\begin{equation}
\label{eq:Caexp}
C_{a,{\rm f}}(x,\alpha_{\rm s}) \;\: = \;\: 
\delta(1-x) \:+\: 
a_{\rm s}\, c_{a,{\rm f}}^{(1)}(x) \:+\: 
a_{\rm s}^{\,2} \, c_{a,{\rm f}}^{(2)}(x) \:+\:
a_{\rm s}^{\,3} \, c_{a,{\rm f}}^{(3)}(x) \:+\:
a_{\rm s}^{\,4} \, c_{a,{\rm f}}^{(4)}(x) \:+\:
\ldots \:\: ,
\end{equation}
and are completely known up to N$^3$LO~\cite{Vermaseren:2005qc,Moch:2008fj}, 
i.e. all terms $c_{a,{\rm f}}^{(3)}$.
At four-loop order a low number of fixed Mellin moments, defined as 
\begin{equation}
\label{eq:mellinN}
  c(N) \; = \; \int_0^1 \! dx \, x^{\,N-1}\, c(x)\, ,
\end{equation}
are available~\cite{Ruijl:2016pkm} as well as the complete soft corrections 
in the limit $x \to 1$ using threshold resummation and QCD factorization in
$d$-dimensions~\cite{Das:2019btv}.

The scale dependence of the PDFs is governed by the well-known evolution equations 
\begin{eqnarray}
\label{eq:evolS}
  \frac{d}{d \ln \mu^2}\, \phi_{\rm f}
  \,=\, \sum_{\rm f^\prime = q,\,\bar{q},\,g}\, 
    P_{\rm ff^\prime}(\alpha_s(\mu^2)) 
    \,\otimes\, 
    \phi_{\rm f^\prime}(x,\mu^2) 
    \, .
\end{eqnarray}
For QCD with $n_f$ quark flavors and with $'\otimes'$ denoting the standard convolution
these are commonly expressed in terms of $2n_f-1$ scalar equations in the flavor non-singlet case, 
and a coupled set of $2 \times 2$ matrix equations in the flavor singlet case.
The evolution kernels, i.e. the space-like splitting functions $P_{\rm ff^\prime}$ 
are calculable in perturbative QCD as well,
\begin{equation}
\label{eq:Pexp}
P_{\rm ff^\prime}(x,\alpha_{\rm s}) \;\: = \;\: 
a_{\rm s} P^{(0)}_{\rm ff^\prime}(x) 
+ a_{\rm s}^2 P^{(1)}_{\rm ff^\prime}(x) 
+ a_{\rm s}^3 P^{(2)}_{\rm ff^\prime}(x) 
+ a_{\rm s}^4 P^{(3)}_{\rm ff^\prime}(x) 
+ a_{\rm s}^5 P^{(4)}_{\rm ff^\prime}(x) 
+ \dots \:\: .
\end{equation}
The NNLO results $P^{(2)}_{\rm ff^\prime}$ are all known~\cite{Moch:2004pa,Vogt:2004mw}. 
At N$^3$LO, i.e., at four loops, the non-singlet quark-quark
splitting functions have been computed in the large-$N_c$ limit and number of Mellin moments for the remaining 
color coefficients are known~\cite{Moch:2017uml,Moch:2018wjh} for a general $SU(N_c)$ gauge theory.
In the flavor-singlet sector the leading large-$n_f$ terms and those 
proportional to quartic color Casimirs are known~\cite{Davies:2016jie,Moch:2018wjh}.
Beyond this order, even some low-$N$ Mellin moments of 
the five-loop contributions to the non-singlet quark-quark splitting function 
$P^{(4)}_{\rm ns}$ have been determined~\cite{Herzog:2018kwj}.

In the following, we give a brief overview of the computational set-up and
work-flow underlying the computations at four loops and beyond.

\subsection{Computational work-flow}

Using the operator product expansion in DIS 
one can relate the product of currents $J_\mu$ in the hadronic tensor in Eq.~\eqref{eq:htensor} 
to Mellin moments of the structure functions $F_{\,2,\:3\:,L}$, see, e.g. \cite{Buras:1979yt}.
The latter, parameterizing the (semi)-inclusive cross section, are then
obtained with the help of the optical theorem 
from the imaginary part of the forward Compton amplitude for the gauge boson-nucleon scattering.
Thus, the computation of QCD corrections in DIS starts from the 
forward Compton amplitude of the corresponding gauge boson-parton scattering
process, using the kinematics of Eq.~\eqref{eq:xb}, 
which gives access to both, the coefficient functions in Eq.~\eqref{eq:Caexp}
and the splitting functions in Eq.~\eqref{eq:Pexp}. 
In case, one is only interested in the latter, the direct computation of operator matrix
elements, evaluated in parton two-point functions, proves more efficient and 
allows for the determination of the anomalous dimensions $\gamma(N)$, i.e., the Mellin transforms 
of the splitting functions, cf. Eq.~\eqref{eq:mellinN}.

\begin{figure}[t]
\sidecaption
\includegraphics[scale=.80]{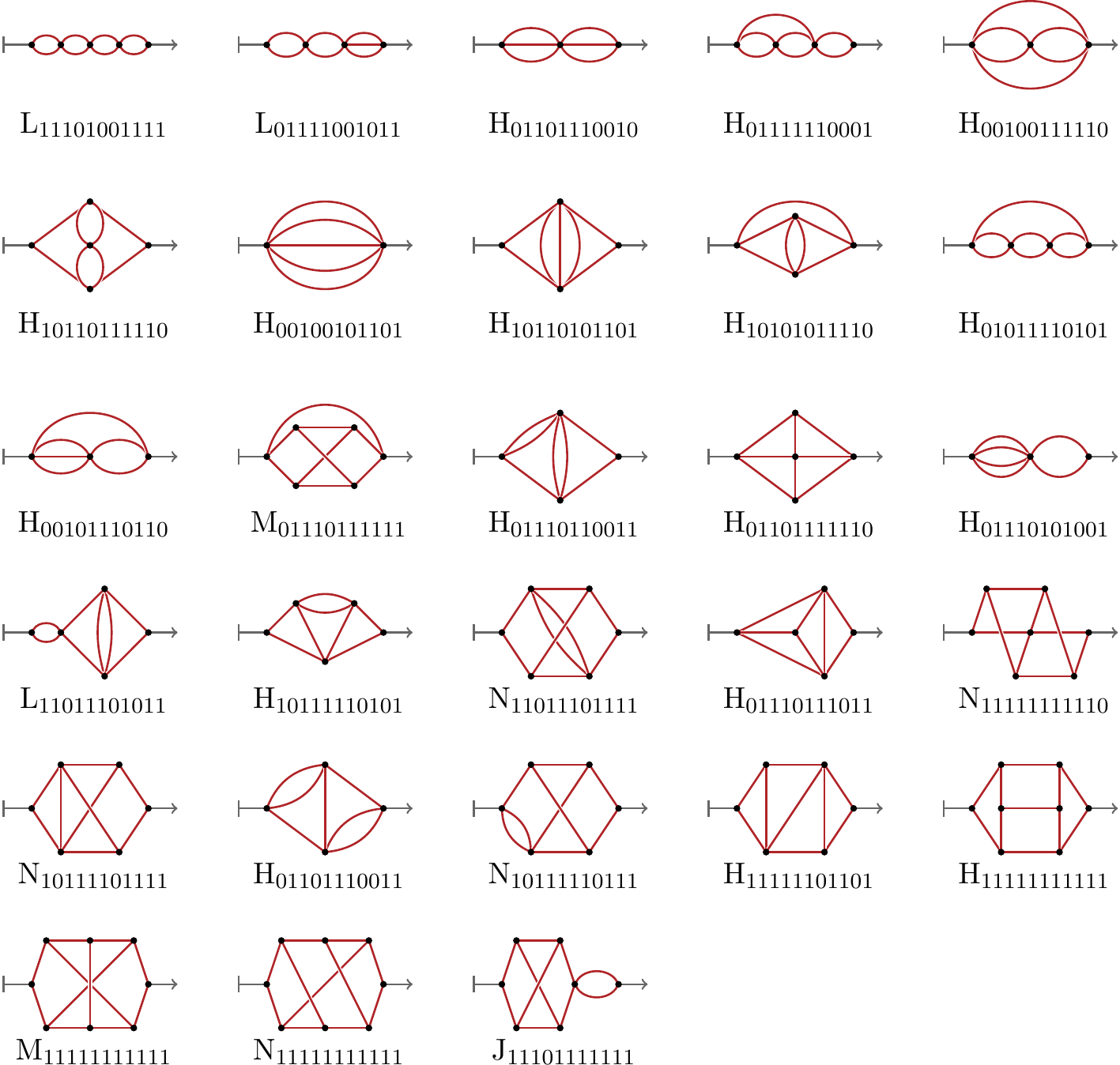}
\caption{Master integrals for four-loop propagators (figure from~\cite{Magerya:2019cvz}).}
\label{fig:virtual}
\end{figure}

The required Feynman diagrams up to four loops can generated using 
the diagram generator {\tt Qgraf}~\cite{Nogueira:1991ex} and 
the group theory factors for a general color $SU(N_c)$ gauge theory can be obtained
with algorithms described in~\cite{vanRitbergen:1998pn}.
The loop integrals are considered in dimensional regularization~\cite{tHooft:1972tcz,Bollini:1972ui},
$d=4-2\varepsilon$, which is the standard framework in perturbative QCD at
higher orders and the integral reductions are performed by means of integration-by-parts
identities (IBP)~\cite{Tkachov:1981wb,Chetyrkin:1981qh}.
The solution of the IBP reductions are encoded in the program {\tt Forcer}~\cite{Ruijl:2017cxj}, which 
performs a parametric reduction of four-loop massless propagator diagrams to master integrals.
The latter are shown in Fig.~\ref{fig:virtual} and their analytic expressions
as a Laurent series in $\varepsilon$ have been computed in Refs.~\cite{Baikov:2010hf,Lee:2011jt}.
The symbolic manipulations employ the computer algebra system 
{\tt Form}~\cite{Vermaseren:2000nd,Kuipers:2012rf,Ruijl:2017dtg} and its
multi-threaded version {\tt TForm}~\cite{Tentyukov:2007mu} in order  
to handle both, the run times and the size of the intermediate expressions
occurring in the reduction of diagrams with high Mellin moments $N$.

The approach delivers results for fixed Mellin moments of the anomalous
dimensions and DIS coefficient functions. 
When enough fixed Mellin moments are available, one can follow the approach of~\cite{Velizhanin:2012nm}, 
and attempt the reconstruction of an analytic expression as a function of $N$ 
in terms of harmonic sums~\cite{Vermaseren:1998uu,Blumlein:1998if}.
In the planar limit, i.e., for large $N_c$, the exact four-loop results for moments up to $N = 20$ are 
sufficient to determine the analytic expressions of the non-singlet quark-quark
anomalous dimensions $\gamma^{(3)}_{\rm ns}(N)$ as a function of $N$ by {\tt LLL}-based 
techniques~\cite{Lenstra1982,axbAlg,DBLP:journals/dcc/Silverman00,Calc} 
and solving systems of Diophantine equations, cf.~\cite{Moch:2017uml} for details.

The bottleneck of the approach via fixed Mellin moments is caused by the high
powers of propagators, which need to undergo the parametric reduction with 
the program {\tt Forcer}~\cite{Ruijl:2017cxj}.
This leads to large intermediate expressions of the order of TByte 
and to long run times of the computer algebra system {\tt Form}.
For example, the computation of the Mellin moment $N=10$ of the quark coefficient function 
in the projection on $F_{L}$ in Eq.~\eqref{eq:FdispQCD} at four loops requires the
evaluation ${\cal O}(3200)$ diagrams with a total of ${\cal O}(800000)$ hrs
CPU time, i.e. almost 100 years altogether. 
Fortunately, the multi-threaded version {\tt TForm} delivers an 
average speed-up factor of ${\cal O}(10)$ and with a cluster of sufficiently
many servers, the problem is doable within half a year of "wall time".

Extensions to five-loop low-$N$ Mellin moments of 
the non-singlet anomalous dimension $\gamma^{(4)}_{\rm ns}(N)$, i.e. the Mellin transform of 
$P^{(4)}_{\rm ns}$, as achieved in~\cite{Herzog:2018kwj}, 
require the computation of five-loop self-energy integrals, which can be 
accomplished with an implementation \cite{Herzog:2017bjx} 
of the local R$^\ast$ operation~\cite{Chetyrkin:1982nn,Chetyrkin:1984xa,Chetyrkin:2017ppe}. 
This allows for the reduction to four-loop integrals, that can be evaluated
again by the {\tt Forcer} program~\cite{Ruijl:2017cxj}. 
However, the size of intermediate expressions and the run times of {\tt Form}
become prohibitively large beyond the fixed values $N=2$ and $N=3$.

\section{Time-like kinematics}
\label{sec:time}

Semi-inclusive $e^+e^-$ annihilation via a virtual photon or $Z$-boson 
with time-like momentum $q$ proceeds as
\begin{eqnarray}
  \label{eq:epem}
  e^{-} \:+\: e^{+} \:\:\rightarrow\:\: \gamma/Z(q) 
  \:\:\rightarrow\:\: h(p) \:+\: X 
  \:\: ,
\end{eqnarray}
where $h(p)$ stands for a specific species of identified hadrons in the final state. 
The time-like kinematics are characterized by the momentum transfer $Q^2 \equiv q^2 > 0$ 
and the Feynman variable is 
\begin{eqnarray}
  \label{eq:xf}
  x_F\,=\, \frac{2\,p\cdot q}{Q^2}\, ,
\end{eqnarray}
with $0 < x_F \leq 1$. In the center-of-mass frame $x_F$ 
is the fraction of the beam energy carried by the hadron $h$.
The space- and time-like processes~\eqref{eq:dis} and~\eqref{eq:epem} are
related by crossing which implies a mapping $x_B \to x_F$ for 
the kinematics in Eqs.~\eqref{eq:xb} and \eqref{eq:xf} and the use of analytic continuation.

In perturbative QCD, the total (angle-integrated) fragmentation function
\begin{equation}
\label{eq:sigmaFtot} 
  \frac{1}{\sigma_{\rm tot}} \: \frac{d \sigma^{\,h}}{dx}
   \;\: = \;\: F^{\:\! h}(x,Q^2) \:\: ,
\end{equation}
as well as the transverse ($F_T^{\:\! h}$), longitudinal ($F_L^{\:\! h}$) and asymmetric ($F_A^{\:\! h}$) 
ones parameterizing the double-differential cross section 
$\,d\sigma^{\,h}/ dx\, d \cos\theta_h$ \cite{Nason:1993xx}, are given by
\begin{equation}
\label{eq:FapQCD}
  F_a^{\:\! h}(x,Q^2) \;\: = \; \sum_{\rm f = q,\,\bar{q},\,g} \;
  \int_x^1 {dz \over z} \; C^{T}_{a,{\rm f}} \left( z,\alpha_{\rm s} (Q^2) \right)
  \,  D_{\rm f}^{\,h} \bigg( \,{x \over z},Q^2\bigg)
  \; + \; {\cal O} \!\left( \frac{1}{Q} \right) 
  \:\: ,
\end{equation}
in terms of the parton fragmentation functions (FFs) $D_{\rm f}^{\,h}$ 
and the (time-like) coefficient functions $C^{T}_{a,\rm f\,}$,
\begin{equation}
\label{eq:Cqexp}
  C^{T}_{a,\rm f} (x,\alpha_{\rm s}) \;\: = \;\: \sigma_{\rm ew} \big(  
  \delta(1-x) \:+\: 
  a_{\rm s}\, c_{a,\rm f}^{T,(1)}(x) \:+\: 
  a_{\rm s}^{\,2} \, c_{a,\rm f}^{T,(2)}(x) \:+\:
  a_{\rm s}^{\,3} \, c_{a,\rm f}^{T,(3)}(x) \:+\:
  \ldots \big) \:\: ,
\end{equation}
where $\sigma_{\rm ew}$ denotes the electroweak pre-factors~\cite{Nason:1993xx}
and the second-order coefficient functions 
are known~\cite{Rijken:1996vr,Rijken:1996ns,Rijken:1996npa,Mitov:2006ic,Mitov:2006wy}, 
while the three-loop corrections $c_{a}^{T,(3)}(x)$ have not been derived so far.

The parton FFs $D_{\rm f}^{\,h}$ obey evolution equations 
analogous to the PDFs in Eq.~\eqref{eq:evolS}, 
\begin{eqnarray}
  \frac{d}{d \ln \mu^2}\, D_{\rm f}^{\,h}
  \,=\, \sum_{\rm f^\prime = q,\,\bar{q},\,g}\, 
    P^{T}_{\rm f^\prime f}(\alpha_s(\mu^2)) 
    \,\otimes\, 
    D_{\rm f^\prime}^{\,h}(x,\mu^2) 
    \, ,
\end{eqnarray}
with time-like splitting functions, but with 
$2 \times 2$ matrix $P^{T}_{\rm f^\prime f}$ in the flavor singlet case 
transposed compared to PDFs in Eq.~\eqref{eq:evolS}.
In perturbative QCD the time-like splitting functions 
can be expanded as 
\begin{equation}
\label{eq:PTexp}
P^{T}_{\rm f^\prime f}(x,\alpha_{\rm s}) \;\: = \;\: 
a_{\rm s} P^{T,(0)}_{\rm f^\prime f}(x) 
+ a_{\rm s}^2 P^{T,(1)}_{\rm f^\prime f}(x) 
+ a_{\rm s}^3 P^{T,(2)}_{\rm f^\prime f}(x) 
+ a_{\rm s}^4 P^{T,(3)}_{\rm f^\prime f}(x) 
+ \dots \:\: ,
\end{equation}
where the NNLO results $P^{T,(2)}_{\rm f^\prime f}$ are all
known~\cite{Mitov:2006ic,Moch:2007tx,Almasy:2011eq,Chen:2020uvt}, while 
at N$^3$LO only the non-singlet quark-quark splitting functions are 
available in the large-$N_c$ limit~\cite{Moch:2017uml}.
These results are all based on analytic continuation from space- to time-like
kinematics and on exploiting reciprocity relations for collinear splitting
functions in QCD.
In the sequel, we discuss the computational work-flow for the direct
calculation of QCD corrections to semi-inclusive $e^+e^-$ annihilation~\eqref{eq:epem}.

\subsection{Inclusive cross-sections}

A practical indirect way of calculating total cross-sections
is the optical theorem. Through it, $\mathcal{O}(\alpha_s^3)$
corrections for $e^+e^-$ annihilation in Eq.~\eqref{eq:epem}
can be expressed in terms of the four-loop propagator
diagrams. In the massless case all the~22 master integrals for
these propagators shown in Fig.~\ref{fig:virtual} 
have been calculated in~\cite{Baikov:2010hf,Lee:2011jt}. 

The direct way on the other hand requires the calculation of all
squared amplitudes with 2, 3, 4, and 5 particles in the final
state (with 3, 2, 1, and 0 loops respectively), e.g.
\begin{equation}
    \sigma \sim \sum_n \int \mathrm{d} \mathrm{PS}_n \left|\langle p_{1},\dots,p_{n}\vert iT\vert q\rangle\right|^2 =
        \int \mathrm{d} \mathrm{PS}_3 \left|
        \vcenter{\hbox{\includegraphics[scale=0.5]{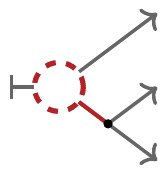}}} +
        \vcenter{\hbox{\includegraphics[scale=0.5]{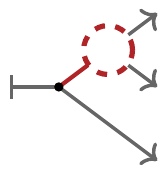}}} + \ldots
        \right|^2 + \dots,
\end{equation}
and integration of those over the respective phase space,
\begin{equation}
    \mathrm{d} \mathrm{PS}_n \equiv
        \left(\prod_{i=1}^{n}\frac{\mathrm{d}^{d}p_{i}}{\left(2\pi\right)^{d-1}} \delta^{+}(p_{i}^{2})\right)
        \left(2\pi\right)^{d} \delta^{d}(q-\sum_{j=1}^{n}p_{j})\, .
    \label{eq:dps}
\end{equation}
Performing the phase space integration analytically quickly turns
out to be the bottleneck: the parameterization of 4- and 5-
particle phase spaces necessarily requires the introduction of square
roots into the integrand, preventing an analytic solution, 
see, e.g., the "tripole parameterization" of the 4-particle phase space in~\cite{Gehrmann-DeRidder:2003pne}.
Instead one should consider both loop and phase space integration appearing in the squared amplitude
together, as a single "cut" diagram:
\begin{equation}
    \int \mathrm{d} \mathrm{PS}_{3}\,
        \vcenter{\hbox{\includegraphics[scale=0.5]{Pics/amp1}}}
        \left(
            \vcenter{\hbox{\includegraphics[scale=0.5]{Pics/amp2}}}
        \right)^{*}
    \equiv
    \int d\mathrm{PS}_{3}\,
        \vcenter{\hbox{\includegraphics[scale=0.5]{Pics/amp1}}}
        \vcenter{\hbox{\includegraphics[scale=0.5]{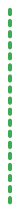}}}
        \vcenter{\hbox{\includegraphics[scale=0.5]{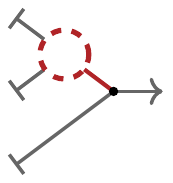}}}
    \equiv
    \vcenter{\hbox{\includegraphics[scale=0.5]{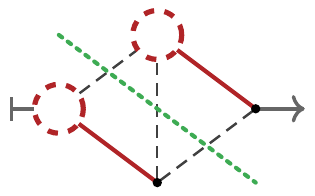}}}\, .
\end{equation}
Then, by applying the idea of "reverse unitarity"~\cite{Anastasiou:2002yz}:
that is, replacing on-shell conditions for final state particles
in Eq.~\eqref{eq:dps} by denominators,
\begin{equation}
    \delta^{+}(p^2) = \frac{1}{2 \pi i} \left( \frac{1}{p^2-i0} - \frac{1}{p^2+i0} \right)\, ,
    \label{eq:delta-prop}
\end{equation}
one can treat each outgoing line as a "cut propagator", and
thus construct IBP relations for cut diagrams.

\begin{figure}[b]
\sidecaption
\includegraphics[scale=.80]{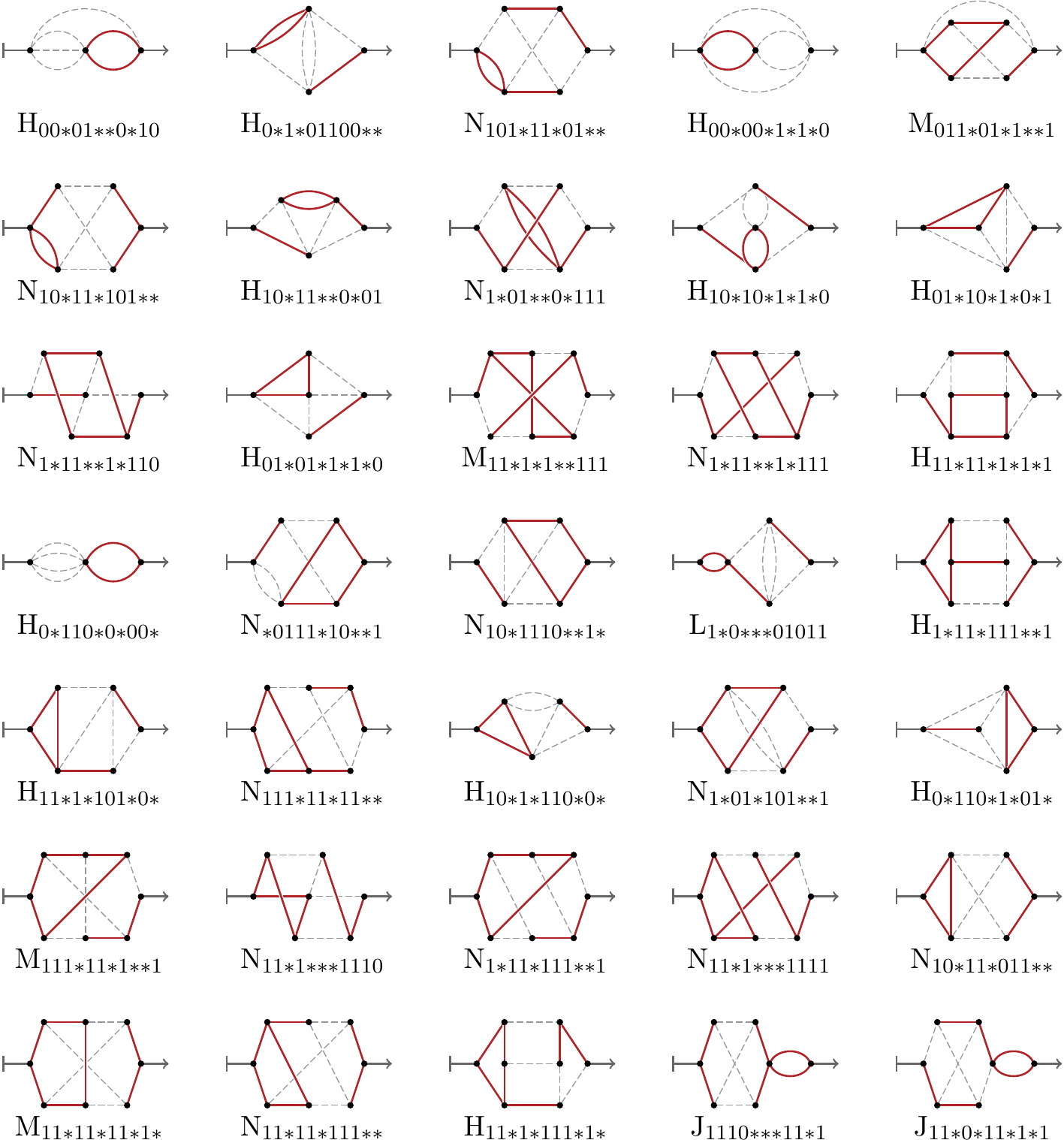}
\caption{Master integrals for all four-particle cuts of four-loop propagators 
(figure from~\cite{Magerya:2019cvz}).}
\label{fig:4pcuts}
\end{figure}

In this way the direct calculation is reduced to calculating the
master integrals for 2-, 3-, 4-, and 5- particle cuts of four-loop
propagators. For this task conventional IBP software can be largely
reused with two modifications: first, any cut propagator raised
to a non-negative power can be set to zero (because $x\,\delta(x)=0$),
and second, when symmetries between diagrams are constructed,
cut propagators should not be symmetrized with the regular ones.

The full set of (massless) master integrals for 5-particle cuts
of four-loop propagators has been first calculated in~\cite{Gituliar:2018bcr},
for 4- and 3-particle cuts in~\cite{Magerya:2019cvz}, and for 2-particle
cuts in~\cite{Heinrich:2007at,Heinrich:2009be,Lee:2010cga}. 
As an example, the set of master integrals for 4-particle cuts is shown in Fig.~\ref{fig:4pcuts}.
Because these integrals are single-scale, it is convenient to calculate them by solving
dimensional recurrence relations (DRR)~\cite{Tarasov:1996br,Tarasov:2000sf}, which
relate the values of these integrals at different values of the
space-time dimension $d$:
\begin{equation}
    I_i(d+2) = M_{ii} I_i(d) + \sum_{j \ne i} M_{ij} I_j(d)\, .
\end{equation}
As explained in~\cite{Magerya:2019cvz}, there is at most a single master
integral per sector, and thus the matrix $M_{ij}$ is triangular. With
the help of the "dimensional recurrence and analyticity"
method of~\cite{Lee:2009dh}, an ansatz for the full solution can be
constructed, leaving only a number of constants undetermined.
Once enough extra information is gathered to fix these constants
(i.e. values of the leading pole coefficients, or several terms
of the $\varepsilon$-expansion computed by alternative means),
{\tt DREAM}~\cite{Lee:2017ftw} can be used to evaluate $I_i(4-2\varepsilon)$ as
a series in $\varepsilon$ with arbitrary precision (thousands
of digits), and these numerical values can then be turned into
analytic expressions in terms of multiple zeta values~\cite{Blumlein:2009cf}
with the help of an integer relation reconstruction algorithm like
{\tt PSQL}~\cite{FergusonBA99}.

The optical theorem (or rather Cutkosky
rules~\cite{Cutkosky:1960sp,tHooft:1973wag}), being the alternative way of
computing fully inclusive quantities, provides an essential
cross-check on these master integrals: the imaginary part of each
four-loop propagator diagram must be equal to a combination of its
cuts. For example:
\begin{equation}
    2\,\mathrm{Im}\,
    \vcenter{\hbox{\includegraphics[scale=0.5]{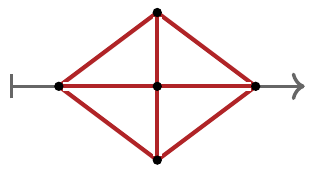}}} =2\,\mathrm{Re}\,
    \vcenter{\hbox{\includegraphics[scale=0.5]{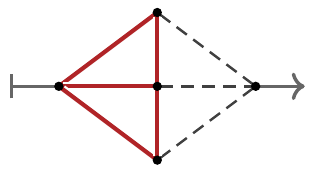}}} +4\,\mathrm{Im}\,
    \vcenter{\hbox{\includegraphics[scale=0.5]{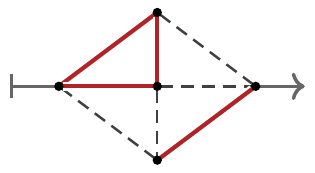}}} -2\,
    \vcenter{\hbox{\includegraphics[scale=0.5]{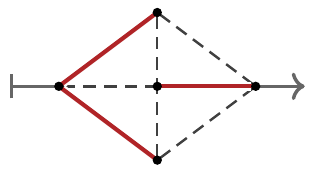}}}\, ,
\end{equation}
where the dashed lines indicate the propagators cut according to
Eq.~\eqref{eq:delta-prop}, see also Fig.~\ref{fig:4pcuts}.

\subsection{Semi-inclusive cross-sections}

Integrals for semi-inclusive cross-sections differ from the
inclusive case by the presence of the scaling parameter $x$
(i.e., $x_F$ in Eq.~\eqref{eq:xf}) 
in the integration measure,
\begin{equation}
  \label{eq:x-cut}
    \mathrm{d} \mathrm{PS}_n(x) \equiv \mathrm{d} \mathrm{PS}_n \, \delta(x-2q\!\cdot\!p_1/q^2),
\end{equation}
so that the semi-inclusive cut diagrams now have the form of
\begin{equation}
    \int \mathrm{d} \mathrm{PS}_3(x) \,
        \vcenter{\hbox{\includegraphics[scale=0.5]{Pics/amp1}}}
        \left(
            \vcenter{\hbox{\includegraphics[scale=0.5]{Pics/amp2}}}
        \right)^{*}
    \equiv
        \vcenter{\hbox{\includegraphics[scale=0.5]{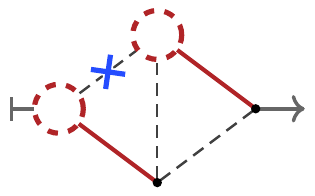}}}\, ,
\end{equation}
where the crossed dashed line corresponds to the constraint in Eq.~\eqref{eq:x-cut}.
The inclusion of $x$ still allows for an IBP reduction if one 
applies Eq.~\eqref{eq:delta-prop} to transform this additional
$\delta$-function into a cut propagator, this time a massive one.
This complicates calculations:
\begin{itemize}
    \item
        first, by introducing linear dependencies between
        denominators of a given diagram; these need to be split
        through partial fractioning, with the end result that a
        single Feynman diagram can now contribute terms to
        several different meta topologies;
    \item
        second, by the increased number of master integrals:
        there are 693 semi-inclusive master integrals (298
        for 5-particle cuts, 277 for 4-particle cuts, 96 for
        3-particle cuts, and 22 for 2-particle cuts) vs. 115
        for the inclusive case;
    \item
        third, by the increase in the size of IBP expressions,
        and the increased computational requirements of the
        reduction;
    \item
        and finally by the fact that one can not easily solve DRR
        for these integrals: the method used for the inclusive
        case largely relied on the numerical evaluation, and having
        a free parameter $x$ makes that impractical (if not
        impossible).
\end{itemize}

The master integrals for semi-inclusive cuts for three-loop propagators
were completed in~\cite{Gituliar:2015iyq,Gituliar:2015pra} and for four-loop
propagators in~\cite{Magerya:thesis}.

A convenient way to calculate the values of these master integrals is
the method of differential equations~\cite{Kotikov:1990kg,Kotikov:1991pm}:
the integrands of cut master integrals can be differentiated
with respect to~$x$, and the obtained expressions can then be reduced back to the
same integrals via IBP relations, resulting in systems of differential
equations of the form
\begin{equation}
    \frac{\partial}{\partial x} I_i(d,x) = M_{ij}(d,x) I_j(d,x)\, .
    \label{eq:diffeq}
\end{equation}
To solve these equations, one follows the observation from~\cite{Henn:2013pwa}:
if there is a basis transformation
\begin{equation}
    J_i(d,x) = T_{ij}(d,x) I_j(d,x)\, ,
    \label{eq:eps-t}
\end{equation}
such that once substituted into Eq.~\eqref{eq:diffeq} factorizes
the dependence of $M$ on $d=4-2\varepsilon$, transforming the
equation into an $\varepsilon$-form,
\begin{equation}
    \frac{\partial}{\partial x} J_i(d,x) = \varepsilon S_{ij}(x) J_j(d,x)\, ,
\end{equation}
then the solution can easily found as a series in $\varepsilon$,
\begin{equation}
    I_i(4-2\varepsilon,x) \equiv \sum_k \varepsilon^k I_i^{k}(x), \qquad
    I_i^{(k)}(x) = \int \mathrm{d}x\,S_{ij}(x) I_j^{(k-1)}(x) + C_i^{(k)}\, .
\end{equation}
Only two issues remain: how to find the transformation matrix
$T_{ij}$ from Eq.~\eqref{eq:eps-t}, and how to fix the integration
constants~$C_i^{(k)}$.

A general algorithm of constructing $\varepsilon$-form transformations
directly from the matrix $M_{ij}$ was presented in~\cite{Lee:2014ioa}
and improved upon in \cite{Blondel:2018mad,Lee:2017oca}. We rely upon the public
implementation of this algorithm, {\tt Fuchsia}~\cite{Gituliar:2017vzm,Gituliar:2016vfa}, 
to find $T_{ij}$, specifically on the new version available at~\cite{Fuchsia}.

To fix the integration constants observe that if one integrates
a semi-inclusive integrals over all $x$, the result should be a
fully inclusive integral. So by writing down equations of the form
\begin{equation}
    \int \mathrm{d}x \,
    \vcenter{\hbox{\includegraphics[scale=0.5]{Pics/amp1x2x}}}
    =
    \vcenter{\hbox{\includegraphics[scale=0.5]{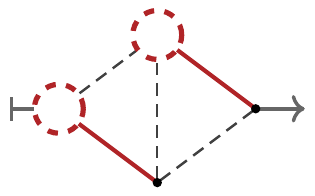}}}
    \, ,
\end{equation}
for each master integral, and inserting the series' in $\varepsilon$
for both the semi-inclusive and the (known) inclusive integrals, all
$C_i^{(k)}$ can be recovered. The only complication here is that
the solution for $I_i(d,x)$ may contain terms $\sim 1/x$, which
would make the integral on the left-hand side divergent if taken
order-by-order in the series. This can be side-stepped by
multiplying the integrand on the left-hand side by $x^m$ with
high enough $m$, and inserting a denominator of the form
$\left(2q\!\cdot\!p_1/q^2\right)^m$ into the diagram on the
right-hand side.

At this stage, it remains to apply the IBP reductions for the semi-inclusive
case to the Feynman diagrams of the individual parton processes 
contributing to the semi-inclusive $e^+e^-$ annihilation in Eq.~\eqref{eq:epem}. 
This will check the NNLO results for the time-like splitting functions $P^{T,(2)}_{\rm f^\prime f}(x)$ 
in Eq.~\eqref{eq:PTexp} by a direct computation and determine the hitherto unknown 
three-loop corrections $c_{a}^{T,(3)}(x)$ in Eq.~\eqref{eq:Cqexp}.

\section{Conclusions}
\label{sec:concl}

The push towards N$^3$LO accuracy in QCD for DIS structure functions or fragmentation functions 
in $e^+e^-$ annihilation requires calculations at four-loop order. 
The efforts are realized with a largely automated work-flow for the generation
of all Feynman diagrams, the parametric IBP reduction to master integrals 
of loop and phase space integrals, for the latter after a mapping with "reverse unitarity"
to loop integrals with cuts, and the computation of the master integrals with
various algorithms, such as DRR or differential equations.
The complexity of the computations, i.e., the size of the expressions, the run
times for IBP reductions and the algorithms for the solution of master integrals
poses challenges to currently available 
computer algebra programs and requires continuous improvements. 
We have presented a brief review of the current status, listing available
results as well as indicating the needs for future improvements.

\section*{Acknowledgments} 
We acknowledge support by Deutsche Forschungsgemeinschaft (DFG) through the
Research Unit FOR 2926, ``Next Generation pQCD for Hadron Structure: Preparing
for the EIC'', project MO 1801/5-1.

%\bibliographystyle{spphys.bst}
%\bibliographystyle{h-physrev3.bst}
%\bibliography{4loopbib}

\begin{thebibliography}{10}

\bibitem{Moch:2004pa}
S.~Moch, J.~A.~M. Vermaseren, and A.~Vogt,
\newblock Nucl. Phys. B {\bf 688}, 101 (2004), hep-ph/0403192.

\bibitem{Vogt:2004mw}
A.~Vogt, S.~Moch, and J.~A.~M. Vermaseren,
\newblock Nucl. Phys. B {\bf 691}, 129 (2004), hep-ph/0404111.

\bibitem{Mitov:2006ic}
A.~Mitov, S.~Moch, and A.~Vogt,
\newblock Phys. Lett. B {\bf 638}, 61 (2006), hep-ph/0604053.

\bibitem{Moch:2007tx}
S.~Moch and A.~Vogt,
\newblock Phys. Lett. B {\bf 659}, 290 (2008), 0709.3899.

\bibitem{Almasy:2011eq}
A.~A. Almasy, S.~Moch, and A.~Vogt,
\newblock Nucl. Phys. B {\bf 854}, 133 (2012), 1107.2263.

\bibitem{Chen:2020uvt}
H.~Chen, T.-Z. Yang, H.~X. Zhu, and Y.~J. Zhu,
\newblock Chin. Phys. C {\bf 45}, 043101 (2021), 2006.10534.

\bibitem{vanNeerven:1991nn}
W.~L. van Neerven and E.~B. Zijlstra,
\newblock Phys. Lett. B {\bf 272}, 127 (1991).

\bibitem{Zijlstra:1991qc}
E.~B. Zijlstra and W.~L. van Neerven,
\newblock Phys. Lett. B {\bf 273}, 476 (1991).

\bibitem{Zijlstra:1992qd}
E.~B. Zijlstra and W.~L. van Neerven,
\newblock Nucl. Phys. B {\bf 383}, 525 (1992).

\bibitem{Moch:1999eb}
S.~Moch and J.~A.~M. Vermaseren,
\newblock Nucl. Phys. B {\bf 573}, 853 (2000), hep-ph/9912355.

\bibitem{Rijken:1996vr}
P.~J. Rijken and W.~L. van Neerven,
\newblock Phys. Lett. B {\bf 386}, 422 (1996), hep-ph/9604436.

\bibitem{Rijken:1996ns}
P.~J. Rijken and W.~L. van Neerven,
\newblock Nucl. Phys. B {\bf 487}, 233 (1997), hep-ph/9609377.

\bibitem{Rijken:1996npa}
P.~J. Rijken and W.~L. van Neerven,
\newblock Phys. Lett. B {\bf 392}, 207 (1997), hep-ph/9609379.

\bibitem{Mitov:2006wy}
A.~Mitov and S. Moch,
\newblock Nucl. Phys. B {\bf 751}, 18 (2006), hep-ph/0604160.

\bibitem{Accardi:2016ndt}
A.~Accardi {\em et~al.},
\newblock Eur. Phys. J. C {\bf 76}, 471 (2016), 1603.08906.

\bibitem{Boer:2011fh}
D.~Boer {\em et~al.},
\newblock (2011), 1108.1713.

\bibitem{Accardi:2012qut}
A.~Accardi {\em et~al.},
\newblock Eur. Phys. J. A {\bf 52}, 268 (2016), 1212.1701.

\bibitem{Blondel:2018mad}
A.~Blondel {\em et~al.},
\newblock {Standard model theory for the FCC-ee Tera-Z stage},
\newblock in {\em {Mini Workshop on Precision EW and QCD Calculations for the
  FCC Studies : Methods and Techniques}}, , CERN Yellow Reports: Monographs
  Vol. 3/2019, Geneva, 2018, CERN, 1809.01830.

\bibitem{vanRitbergen:1997va}
T.~van Ritbergen, J.~A.~M. Vermaseren, and S.~A. Larin,
\newblock Phys. Lett. B {\bf 400}, 379 (1997), hep-ph/9701390.

\bibitem{Czakon:2004bu}
M.~Czakon,
\newblock Nucl. Phys. B {\bf 710}, 485 (2005), hep-ph/0411261.

\bibitem{Baikov:2016tgj}
P.~A. Baikov, K.~G. Chetyrkin, and J.~H. K\"uhn,
\newblock Phys. Rev. Lett. {\bf 118}, 082002 (2017), 1606.08659.

\bibitem{Herzog:2017ohr}
F.~Herzog, B.~Ruijl, T.~Ueda, J.~A.~M. Vermaseren, and A.~Vogt,
\newblock JHEP {\bf 02}, 090 (2017), 1701.01404.

\bibitem{Luthe:2017ttg}
T.~Luthe, A.~Maier, P.~Marquard, and Y.~Schr\"oder,
\newblock JHEP {\bf 10}, 166 (2017), 1709.07718.

\bibitem{Vermaseren:2005qc}
J.~A.~M. Vermaseren, A.~Vogt, and S.~Moch,
\newblock Nucl. Phys. B {\bf 724}, 3 (2005), hep-ph/0504242.

\bibitem{Moch:2008fj}
S.~Moch, J.~A.~M. Vermaseren, and A.~Vogt,
\newblock Nucl. Phys. B {\bf 813}, 220 (2009), 0812.4168.

\bibitem{Ruijl:2016pkm}
B.~Ruijl, T.~Ueda, J.~A.~M. Vermaseren, J.~Davies, and A.~Vogt,
\newblock PoS {\bf LL2016}, 071 (2016), 1605.08408.

\bibitem{Das:2019btv}
G.~Das, S. Moch, and A.~Vogt,
\newblock JHEP {\bf 03}, 116 (2020), 1912.12920.

\bibitem{Moch:2017uml}
S.~Moch, B.~Ruijl, T.~Ueda, J.~A.~M. Vermaseren, and A.~Vogt,
\newblock JHEP {\bf 10}, 041 (2017), 1707.08315.

\bibitem{Moch:2018wjh}
S.~Moch, B.~Ruijl, T.~Ueda, J.~A.~M. Vermaseren, and A.~Vogt,
\newblock Phys. Lett. B {\bf 782}, 627 (2018), 1805.09638.

\bibitem{Davies:2016jie}
J.~Davies, A.~Vogt, B.~Ruijl, T.~Ueda, and J.~A.~M. Vermaseren,
\newblock Nucl. Phys. B {\bf 915}, 335 (2017), 1610.07477.

\bibitem{Herzog:2018kwj}
F.~Herzog {\em et~al.},
\newblock Phys. Lett. B {\bf 790}, 436 (2019), 1812.11818.

\bibitem{Buras:1979yt}
A.~J. Buras,
\newblock Rev. Mod. Phys. {\bf 52}, 199 (1980).

\bibitem{Magerya:2019cvz}
V.~Magerya and A.~Pikelner,
\newblock JHEP {\bf 12}, 026 (2019), 1910.07522.

\bibitem{Nogueira:1991ex}
P.~Nogueira,
\newblock J. Comput. Phys. {\bf 105}, 279 (1993).

\bibitem{vanRitbergen:1998pn}
T.~van Ritbergen, A.~N. Schellekens, and J.~A.~M. Vermaseren,
\newblock Int. J. Mod. Phys. A {\bf 14}, 41 (1999), hep-ph/9802376.

\bibitem{tHooft:1972tcz}
G.~'t~Hooft and M.~J.~G. Veltman,
\newblock Nucl. Phys. B {\bf 44}, 189 (1972).

\bibitem{Bollini:1972ui}
C.~G. Bollini and J.~J. Giambiagi,
\newblock Nuovo Cim. B {\bf 12}, 20 (1972).

\bibitem{Tkachov:1981wb}
F.~V. Tkachov,
\newblock Phys. Lett. B {\bf 100}, 65 (1981).

\bibitem{Chetyrkin:1981qh}
K.~G. Chetyrkin and F.~V. Tkachov,
\newblock Nucl. Phys. B {\bf 192}, 159 (1981).

\bibitem{Ruijl:2017cxj}
B.~Ruijl, T.~Ueda, and J.~A.~M. Vermaseren,
\newblock Comput. Phys. Commun. {\bf 253}, 107198 (2020), 1704.06650.

\bibitem{Baikov:2010hf}
P.~A. Baikov and K.~G. Chetyrkin,
\newblock Nucl. Phys. B {\bf 837}, 186 (2010), 1004.1153.

\bibitem{Lee:2011jt}
R.~N. Lee, A.~V. Smirnov, and V.~A. Smirnov,
\newblock Nucl. Phys. B {\bf 856}, 95 (2012), 1108.0732.

\bibitem{Vermaseren:2000nd}
J.~A.~M. Vermaseren,
\newblock (2000), math-ph/0010025.

\bibitem{Kuipers:2012rf}
J.~Kuipers, T.~Ueda, J.~A.~M. Vermaseren, and J.~Vollinga,
\newblock Comput. Phys. Commun. {\bf 184}, 1453 (2013), 1203.6543.

\bibitem{Ruijl:2017dtg}
B.~Ruijl, T.~Ueda, and J.~Vermaseren,
\newblock (2017), 1707.06453.

\bibitem{Tentyukov:2007mu}
M.~Tentyukov and J.~A.~M. Vermaseren,
\newblock Comput. Phys. Commun. {\bf 181}, 1419 (2010), hep-ph/0702279.

\bibitem{Velizhanin:2012nm}
V.~N. Velizhanin,
\newblock Nucl. Phys. B {\bf 864}, 113 (2012), 1203.1022.

\bibitem{Vermaseren:1998uu}
J.~A.~M. Vermaseren,
\newblock Int. J. Mod. Phys. A {\bf 14}, 2037 (1999), hep-ph/9806280.

\bibitem{Blumlein:1998if}
J.~Bl{\"u}mlein and S.~Kurth,
\newblock Phys. Rev. D {\bf 60}, 014018 (1999), hep-ph/9810241.

\bibitem{Lenstra1982}
A.~K. Lenstra, H.~W. Lenstra, and L.~Lov{\'a}sz,
\newblock Mathematische Annalen {\bf 261}, 515 (1982).

\bibitem{axbAlg}
K.~Matthews,
\newblock (unpublished), summarized in \cite{DBLP:journals/dcc/Silverman00};
  see pp. 16/17 .

\bibitem{DBLP:journals/dcc/Silverman00}
J.~H. Silverman,
\newblock Des. Codes Cryptography {\bf 20}, 5 (2000).

\bibitem{Calc}
{\tt http://www.numbertheory.org/calc/krm\_calc.html}.

\bibitem{Herzog:2017bjx}
F.~Herzog and B.~Ruijl,
\newblock JHEP {\bf 05}, 037 (2017), 1703.03776.

\bibitem{Chetyrkin:1982nn}
K.~G. Chetyrkin and F.~V. Tkachov,
\newblock Phys. Lett. B {\bf 114}, 340 (1982).

\bibitem{Chetyrkin:1984xa}
K.~G. Chetyrkin and V.~A. Smirnov,
\newblock Phys. Lett. B {\bf 144}, 419 (1984).

\bibitem{Chetyrkin:2017ppe}
K.~G. Chetyrkin,
\newblock (2017), 1701.08627.

\bibitem{Nason:1993xx}
P.~Nason and B.~R. Webber,
\newblock Nucl. Phys. B {\bf 421}, 473 (1994),
\newblock [Erratum: Nucl.Phys.B 480, 755 (1996)].

\bibitem{Gehrmann-DeRidder:2003pne}
A.~Gehrmann-De~Ridder, T.~Gehrmann, and G.~Heinrich,
\newblock Nucl. Phys. B {\bf 682}, 265 (2004), hep-ph/0311276.

\bibitem{Anastasiou:2002yz}
C.~Anastasiou and K.~Melnikov,
\newblock Nucl. Phys. B {\bf 646}, 220 (2002), hep-ph/0207004.

\bibitem{Gituliar:2018bcr}
O.~Gituliar, V.~Magerya, and A.~Pikelner,
\newblock JHEP {\bf 06}, 099 (2018), 1803.09084.

\bibitem{Heinrich:2007at}
G.~Heinrich, T.~Huber, and D.~Maitre,
\newblock Phys. Lett. B {\bf 662}, 344 (2008), 0711.3590.

\bibitem{Heinrich:2009be}
G.~Heinrich, T.~Huber, D.~A. Kosower, and V.~A. Smirnov,
\newblock Phys. Lett. B {\bf 678}, 359 (2009), 0902.3512.

\bibitem{Lee:2010cga}
R.~N. Lee, A.~V. Smirnov, and V.~A. Smirnov,
\newblock JHEP {\bf 04}, 020 (2010), 1001.2887.

\bibitem{Tarasov:1996br}
O.~V. Tarasov,
\newblock Phys. Rev. D {\bf 54}, 6479 (1996), hep-th/9606018.

\bibitem{Tarasov:2000sf}
O.~V. Tarasov,
\newblock Nucl. Phys. B Proc. Suppl. {\bf 89}, 237 (2000), hep-ph/0102271.

\bibitem{Lee:2009dh}
R.~N. Lee,
\newblock Nucl. Phys. B {\bf 830}, 474 (2010), 0911.0252.

\bibitem{Lee:2017ftw}
R.~N. Lee and K.~T. Mingulov,
\newblock (2017), 1712.05173.

\bibitem{Blumlein:2009cf}
J.~Bl{\"u}mlein, D.~J. Broadhurst, and J.~A.~M. Vermaseren,
\newblock Comput. Phys. Commun. {\bf 181}, 582 (2010), 0907.2557.

\bibitem{FergusonBA99}
H.~R.~P. Ferguson, D.~H. Bailey, and S.~Arno,
\newblock Math. Comput. {\bf 68}, 351 (1999).

\bibitem{Cutkosky:1960sp}
R.~E. Cutkosky,
\newblock J. Math. Phys. {\bf 1}, 429 (1960).

\bibitem{tHooft:1973wag}
G.~'t~Hooft and M.~J.~G. Veltman,
\newblock NATO Sci. Ser. B {\bf 4}, 177 (1974).

\bibitem{Gituliar:2015iyq}
O.~Gituliar,
\newblock JHEP {\bf 02}, 017 (2016), 1512.02045.

\bibitem{Gituliar:2015pra}
O.~Gituliar and S.~Moch,
\newblock Acta Phys. Polon. B {\bf 46}, 1279 (2015), 1505.02901.

\bibitem{Magerya:thesis}
V.~Magerya,
\newblock (2021), Ph.D.thesis (Universit{\"a}t Hamburg).

\bibitem{Kotikov:1990kg}
A.~V. Kotikov,
\newblock Phys. Lett. B {\bf 254}, 158 (1991).

\bibitem{Kotikov:1991pm}
A.~V. Kotikov,
\newblock Phys. Lett. B {\bf 267}, 123 (1991),
\newblock [Erratum: Phys.Lett.B 295, 409--409 (1992)].

\bibitem{Henn:2013pwa}
J.~M. Henn,
\newblock Phys. Rev. Lett. {\bf 110}, 251601 (2013), 1304.1806.

\bibitem{Lee:2014ioa}
R.~N. Lee,
\newblock JHEP {\bf 04}, 108 (2015), 1411.0911.

\bibitem{Lee:2017oca}
R.~N. Lee and A.~A. Pomeransky,
\newblock (2017), 1707.07856.

\bibitem{Gituliar:2017vzm}
O.~Gituliar and V.~Magerya,
\newblock Comput. Phys. Commun. {\bf 219}, 329 (2017), 1701.04269.

\bibitem{Gituliar:2016vfa}
O.~Gituliar and V.~Magerya,
\newblock PoS {\bf LL2016}, 030 (2016), 1607.00759.

\bibitem{Fuchsia}
{\tt https://github.com/magv/fuchsia.cpp}.

\end{thebibliography}
%\input{moch.bbl}

\end{document}